\documentclass[aps,prb,twocolumn,groupedaddress,showpacs]{revtex4}
\usepackage{graphicx}

\begin{document}


\title{Magnetic ground state of CeNi$_{1-x}$Cu$_{x}$: A calorimetric investigation}


\author{N. Marcano, J. I. Espeso, J. C. G\'{o}mez Sal, J. Rodr\'{\i}guez 
Fern\'{a}ndez}
\affiliation{Departamento de Ciencias de la Tierra y F\'{\i}sica de la Materia
Condensada, Facultad de Ciencias, Universidad de Cantabria, 39005 Santander,
Spain.}
\author{J. Herrero-Albillos, F. Bartolom\'{e}}
\affiliation{Instituto de Ciencia de Materiales de Arag\'{o}n, CSIC-Universidad 
de Zaragoza, 50009 Zaragoza, Spain.}



\date{\today}

\begin{abstract}
CeNi$_{1-x}$Cu$_{x}$ is a substitutional magnetic system where the interplay 
of the different magnetic interactions leads to the disappearance 
of the long-range magnetic order on the CeNi side.

The existence of inhomogeneities (spin-clusters or phase coexistence) 
has been previously detected by magnetization and muon spectroscopy 
($\mu$SR) measurements. These inhomogeneities are always observed 
regardless of
the different preparation methods and must, then, be considered 
as intrinsic.

We present a detailed specific heat study in a large temperature 
range of 0.2 K to 300 K. The analysis of these data, considering 
also previous neutron scattering, magnetic characterization and $\mu$SR 
results, allows us to present a convenient description of the system 
as inhomogeneous on the nanometric scale. Two regimes are detected 
in the compositional range depending on the dominant RKKY or 
Kondo interactions. We propose that the long-range magnetic order 
at low temperatures is achieved by a percolative process of magnetic 
clusters that become static below the freezing temperature $T_{f}$. In 
this scenario the existence of a Quantum Critical Point at the 
magnetic-nonmagnetic crossover must be discarded. This situation 
should be considered as an example for other substitutional compounds 
with anomalous magnetic or superconducting properties.
\end{abstract}


\maketitle

\section{Introduction}
Strongly correlated electron metals have been for a long time now one 
of the most fruitful fields for the discovery of new physical phenomena 
(heavy fermions, non-Fermi-liquids (NFL), Quantum Critical Points (QCP) or 
unconventional superconductivity), some of them pointing towards 
a possible new description of the quasiparticles ground state. 
At the same time, the study of such magnetic compounds has provided 
new insight into the understanding of the basic magnetic interactions 
(RKKY, 4f-3d hybridization, crystal electric field, etc.) and 
the interplay between them.

In many of these systems an additional problem appears when we 
are dealing with atomic substitutions. This problem is 
described in general as ``disorder effects'' and could consist 
of compositional local inhomogeneities, phase coexistence, and 
clusterization processes that in most cases are intrinsic 
to the samples and cannot be avoided by means of different preparation 
methods or supplementary thermal treatments.

Specific heat measurements ($C_{P}$) provide a powerful method for studying 
the behavior of these compounds\cite{Gopal} because:
\begin{itemize}
\item[i)] The magnetic contribution to the specific heat and the 
corresponding magnetic entropy are related to 
the energy levels of the magnetic ions (Cerium in our case) and 
the shape of the anomalies corresponding to the magnetic phase 
transition gives a valuable indication as to the nature of the 
transition. 
\item[ii)] The electronic coefficient $\gamma$ provides 
relevant information concerning the conduction band density of 
states at the Fermi level. 
\item[iii)] At higher temperatures the shape 
of the thermal variation of the specific heat depends on the 
crystal electric field splitting. 
\item[iv)] The very low temperature 
$C_{P}/T$ behavior can show departures from the predictions 
of the Fermi-liquid theory,\cite{Coleman} which could serve as 
the sign of the different scenarios devoted to explaining the 
physics involved in these processes (spin fluctuations that exist 
close to a QPC,\cite{Moriya,Coleman99,Schroder} Griffiths 
phase situation,\cite{CastroNeto} disorder-driven mechanisms 
as ``Kondo disorder'',\cite{Bernal,Miranda} etc.). 
\end{itemize}

All these 
features can be used in the analysis of the CeNi$_{1-x}$Cu$_{x}$ system, 
which has been studied over the last few years with different macroscopic 
and microscopic techniques such as magnetization, resistivity, 
neutron diffraction, $\mu$SR spectroscopy, etc. The main features 
of this system at the present stage can be summarized as follows:

\begin{itemize}
\item[a)] A complex magnetic behavior was determined\cite{Espeso} 
with a change from an Antiferromagnetic (AFM) (CeCu) to Ferromagnetic 
(FM) ground state well characterized by neutron diffraction in 
CeNi$_{0.4}$Cu$_{0.6}$ (Ref.~\onlinecite{GomezSal}) evolving 
towards the evanescence of 
the long-range magnetic order around CeNi$_{0.8}$Cu$_{0.2}$.

\item[b)]  A ``spin-glass-like'' phase was surprisingly found at temperatures 
above the ferromagnetic order state.\cite{Soldevilla} A ``cluster-glass'' 
scenario seems the most plausible hypothesis according 
to ac and dc magnetization measurements.\cite{GomezSal02} 

\item[c)]  In fact, recent $\mu$SR studies confirm the presence 
of a highly inhomogeneous magnetic state at temperatures extending 
from the long-range ordered ground state to the paramagnetic 
regime. This inhomogeneous magnetic state consists of long-range 
ordered and nonordered fractions, the latter increasing with 
temperature.\cite{Marcano}

\item[d)]  The Ni-rich part of the diagram is very sensitive to thermal 
treatments due to the proximity to the crystalline structural 
change (FeB to CrB).\cite{Marcano03} There appears a marked 
decrease in magnetization as a consequence of the strong 
reduction of the magnetic Ce moments due to the increasing hybridization 
effects. Some traces of hysteresis however, have been detected 
at very low temperatures in the CeNi$_{0.85}$Cu$_{0.15}$ and 
CeNi$_{0.9}$Cu$_{0.1}$ 
compounds that show that the magnetic moments have not been completely 
exhausted yet.\cite{Marcano04}
\end{itemize}

Many questions still await answers in order for 
there to be a full understanding 
of such complex behavior. The aim of this paper is to show how 
the specific heat measurements performed in these series between 
0.2 K and 300 K provide valuable information that sheds light on 
the underlying physics in the system.

Starting in the first part of the article with a detailed presentation 
of the preparation methods, crystallography, and quality control 
of the samples, we go on to present the specific heat measurements 
and their corresponding analysis. Special care has been taken 
in order to elucidate the nature of the anomalies found in these 
compounds. Particular attention has been paid to the analysis 
of the low temperature regime at the compositional region near 
the magnetic-nonmagnetic crossover.

\section{Experimental details}
\subsection{Instruments}
The specific heat measurements in the range 0.2 K$<$$T$$<$6 K 
were performed at the Instituto de Ciencia de Materiales de Arag\'{o}n 
on a fully automated quasiadiabatic calorimeter\cite{Bartolome} 
refrigerated by adiabatic demagnetization of a paramagnetic 
salt\cite{Miedema,Algra} using the heat pulse technique and germanium 
thermometry over the whole temperature range. Small ingots of around 0.5 g 
weight were used for the measurement, mixed with Apiezon N grease 
to achieve good thermal contact between the sample and the 
calorimetric set (heater and thermometer) even at the lowest 
temperature. The calorimeter is equipped with a mechanical switch that 
allows temperatures down to 0.2 K to be reached. The absolute accuracy 
of the instrument has been estimated to be about 1\%.

 The calorimetric data between 2 K and 300 K were obtained in a 
commercial Quantum Design microcalorimeter at the University 
of Cantabria using the relaxation technique. In this case, thin 
slab shape samples 0.5mm thick and weighing around 8 mg were used. 
Matching between both sets of data in the common temperature 
range is quite satisfactory.

\subsection{\label{Samples}Samples}

The samples in the present work are those with $x$=0.1, 0.15, 
0.2, 0.3, 0.4, 0.5, 0.6, 0.7, 0.8, and 0.9. They are all polycrystalline 
samples prepared by carefully melting together stoichiometric 
amounts of the appropriate high-purity starting elements in an 
arc melting furnace under inert Ar atmosphere. Five melts were 
performed with flipping of the arc-melted button between melts 
in order to improve homogeneity. 

 The crystalline structures of the CeNi$_{1-x}$Cu$_{x}$ compounds studied 
have been determined by x-ray diffraction at 300 K. The alloys 
with $x$$>$0.15 crystallize in the FeB-type orthorhombic structure 
(Pnma space group), whereas the alloys with $x$$\leq$0.15 
crystallize in the CrB- type orthorhombic structure (Cmcm
space group).\cite{Marcano04} Both crystalline structures are 
built from Ce trigonal prisms (with a transition metal in the 
center). Both structures only differ in the relative disposition 
of the trigonal prisms, which is the reason why Ce-Ce and Ce-(Ni,Cu) 
distances evolve continuously from one structure to the 
other.\cite{Soldevilla98}

The Rietveld analyses were performed considering Ni and Cu atoms 
randomly distributed on the same 4c site while the Ce lies on 
the other 4c site.

The narrowness of the x-ray peaks of the as-quenched samples 
spectra guarantees the good crystallization of the samples except 
for the Cu-rich ones; i. e.,  $x$=0.8 and 0.9, which showed 
a broadening of the diffraction peaks. A high vacuum annealing 
at 425 $^\circ$C (just below the melting point of CeCu) for 
one week was carried out on the samples with $x$$>$0.2. Noticeable narrowing 
of the peaks was induced only for $x$=0.8 and 0.9 samples, whereas 
no significant modifications were observed for 
any of the other compositions. 

\begin{table}
    \caption{\label{table1} Crystallographic data of the CeNi$_{1-x}$Cu$_{x}$ 
    compounds obtained at 300 K from x-ray diffraction data.}
    \begin{ruledtabular}
	\begin{tabular}{cccccc}
	    x & Structure & a(\AA) & b(\AA) & c(\AA) & V(\AA$^{3}$)\\
	    \hline
	    0.1 & CrB & 3.800(4) & 10.490(1) & 4.355(7) & 173.7(3)  \\
	    0.15 & CrB & 3.814(3) & 10.565(8) & 4.390(5) & 176.9(3)  \\
	    0.2 & FeB & 7.267(4) & 4.425(2) & 5.598(3) & 180.1(1)  \\
	    0.3 & FeB & 7.280(3) & 4.428(2) & 5.602(2) & 180.6(1)  \\
	    0.4 & FeB & 7.304(2) & 4.445(1) & 5.624(2) & 182.6(1)  \\
	    0.5 & FeB & 7.324(4) & 4.471(2) & 5.646(3) & 184.9(1)  \\
	    0.6 & FeB & 7.354(1) & 4.500(5) & 5.658(5) & 187.2(1)  \\
	    0.7 & FeB & 7.369(4) & 4.506(2) & 5.674(2) & 188.4(1)  \\
	    0.8 & FeB & 7.383(4) & 4.528(2) & 5.688(3) & 190.2(2)  \\
	    0.9 & FeB & 7.410(3) & 4.543(1) & 5.682(2) & 191.3(1)  \\
	\end{tabular}
    \end{ruledtabular}
\end{table}

The compounds with compositions close to the FeB-CrB type structure 
change ($x$=0.1, 0.15, and 0.2), present signs of mixture of FeB/CrB 
phases for the as quenched preparations. The  
stabilization of the proper structure in each case was reached 
with a high vacuum annealing at 440 $^\circ$C for one week.
 
\begin{figure}
    \centerline{\includegraphics[width=7cm]{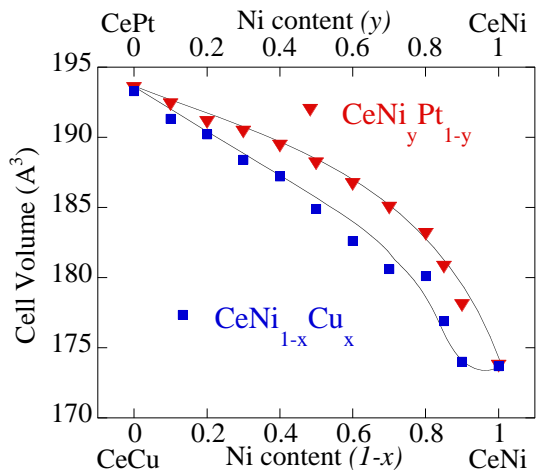}}
    \caption{\label{fig1} Cell volume derived from x-ray diffraction 
    for CeNi$_{1-x}$Cu$_{x}$ together with that for CeNi$_{y}$Pt$_{1-y}$.
    The drawn lines are just a visual guideline.}
\end{figure}

Table~\ref{table1} summarizes the crystallographic data derived from the 
Rietveld refinements at 300 K for the studied compounds. The cell 
volume increases with the Cu content; the evolution is quite 
similar to that reported for CeNi$_{y}$Pt$_{1-y}$;\cite{Gignoux} 
both these sets of data are plotted in Fig.~\ref{fig1}.

In order to get a better characterization of the microstructure 
of the samples, a Scanning Electron Microscopy (SEM) study has 
also been performed.

The intensity of backscattered electron images taken with a SEM 
is proportional to the local atomic number of the sample. Therefore, 
these backscattered electron images yield semiquantitative information 
about composition homogeneity in a sample. Samples were prepared 
for SEM observations by using grinding paper discs of different 
grain sizes and final polishing with diamond paste. The SEM backscattered 
electron images were produced with a JEOL Electron Microscope 
at the University of Cantabria. Zooms between 350 and 2000 were 
reached and quantitative Electron Probe MicroAnalysis techniques 
were used to verify the chemical composition of the phases. We 
used secondary electron images to distinguish composition fluctuations 
from physical voids in the sample by comparing them with the backscattered 
electron images.

The backscattered electron images revealed that most of the samples 
volume consist of a matrix with the expected nominal composition 
in each case. Electron Probe MicroAnalysis measurements revealed 
that this matrix is homogeneous in all the scanned regions 
($\sim$5$\mu$m$^{2}$). 
We observed that as Cu concentration increases, samples tend to 
be very brittle and so, the tendency to present cracks in the 
surface becomes more important.

\section{Results}
\begin{figure}
    \centerline{\includegraphics[width=7cm]{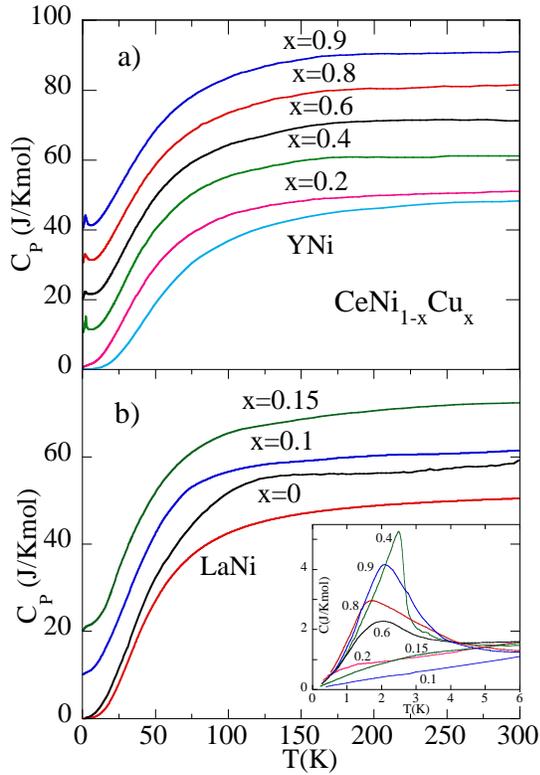}}
    \caption{\label{fig2} a) Temperature dependence of the specific 
    heat for CeNi$_{1-x}$Cu$_{x}$ compounds with $x$$>$0.15 and YNi. 
    The YNi data has been fitted (solid line) considering a Debye and 
    an electronic term (see text for details). b) The specific heat 
    for $x$$\leq$0.15 compounds and LaNi. The CeNi and LaNi data were 
    taken from Ref.~\onlinecite{Gignoux83}. The data corresponding to different 
    compounds have been shifted for clarity. The inset in Fig.~\ref{fig2}b 
    shows the total specific heat in the low temperature regime.}
\end{figure}

The specific heat measurements performed in the temperature 
range 0.2 K to 300 K on the compositions with FeB-type structure 
($x$$>$0.15) and also the isostructural YNi are shown in Fig.~\ref{fig2}a. 
The temperature dependence of $C_{P}$ for YNi can be accounted for
if an electronic term and a lattice one following the 
Debye function are taken into consideration. The fit of 
these data yields the values 
$\gamma$=5 mJ/K$^{2}$ mol and $\theta_{D}$=250 K, which are in good agreement 
with previous results.\cite{Blanco} Fig.~\ref{fig2}b displays the 
temperature dependence of the specific heat for the compositions 
with CrB-type structure ($x$$\leq$0.15) together with the already 
reported data for the specific heat of CeNi and the isostructural 
nonmagnetic LaNi compound.\cite{Gignoux83} In that case, the 
temperature dependence of $C_{P}$ for LaNi follows a Debye function 
with $\gamma$=5 mJ/K$^{2}$ mol and $\theta_{D}$=190 K.\cite{Gignoux83}

Inset of Fig.~\ref{fig2}b shows the low temperature region (from 0.2 K 
to 6 K) of the total specific heat. The origin and nature of the 
anomalies shown here will be explained later on.

\subsection{Estimate of the electronic coefficient $\gamma$}
\begin{figure}
    \centerline{\includegraphics[width=7cm]{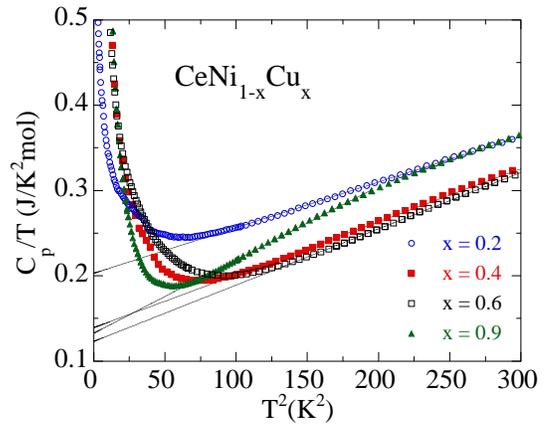}}
    \caption{\label{fig3} Specific heat plotted as $C_{P}/T$ vs. $T^{2}$ 
    showing the temperature range with a linear behavior whose 
    extrapolation to $T$=0 K allows the estimation of the electronic 
    specific heat coefficient $\gamma$. Only some compositions are 
    presented for clarity.}
\end{figure}

\begin{figure}
    \centerline{\includegraphics[width=7cm]{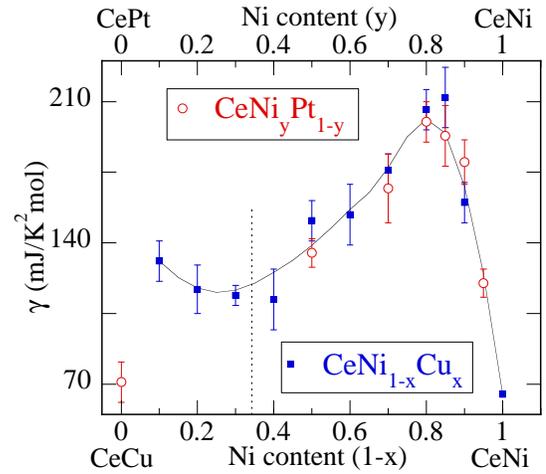}}
    \caption{\label{fig4} Electronic specific-heat coefficient 
    ($\gamma$) for CeNi$_{1-x}$Cu$_{x}$ series as a function of Ni 
    composition ($1-x$) and for CeNi$_{y}$Pt$_{1-y}$ series as a function 
    of Ni ($y$). The solid line is drawn as a guide only. The dashed 
    lines set the crossover between the two compositional 
    regimes defined in the text.}
\end{figure}

The value of the linear electronic coefficient of the specific 
heat ($\gamma$) has been taken as the extrapolation of the 
linear part of the $C_{P}/T$ vs $T^{2}$ curves at zero temperature, 
according to the law $C_{P}$=$\gamma T$+$\beta T^{3}$ followed at 
low temperatures, then overlooking any anomalous 
contribution. The values obtained in this way are indicative 
of the electronic correlation enhancement, i. e., how far the 
corresponding compound is from the free electron picture. Fig.~\ref{fig3} 
illustrates this extrapolation for some selected compositions 
of the series. The differences over the series are clearly seen 
both in the size of the $C_{P}/T$ values and in the slopes of the 
$C_{P}/T$ vs $T^{2}$ data that correspond to the differences in the 
Sommerfeld coefficient ($\gamma$) and the Debye temperature 
($\theta_{D}$) respectively. Table~\ref{table2} reports the values of $\gamma$ 
obtained for all the compositions. According to the reported values 
(130-220 mJ/K$^{2}$ mol) these samples can be considered as moderate 
heavy fermions. The compositional dependence of $\gamma$ is plotted 
in Fig.~\ref{fig4}. We have shown in the same figure the dependence 
of $\gamma$ with composition for CeNi$_{y}$Pt$_{1-y}$,\cite{Blanco94} 
another Ce-system typifying the evolution from a localized magnetic 
state to a delocalized one according to the Doniach diagram.\cite{Doniach} 
In the latter case, the value of the $\gamma$ coefficient increases 
when the Ni content does (increasing hybridization) and reaches 
a maximum value of more than 200 mJ/K$^{2}$ mol around the crossover 
where the change from a magnetic localized regime to a delocalized 
one takes place. Once the 4f-delocalization is predominant, the $\gamma$ 
value decreases with the increasing Ni concentration.

\begin{table*}
    \caption{\label{table2} Magnetic characteristics of the compounds studied: 
    Critical temperature ($T_{C}$);  
    reduced magnetic entropy at $T_{C}$ ($S_{mag}$($T_{C}$)/Rln2); 
    linear electronic specific-heat coefficient ($\gamma$); magnetic moment 
    per Ce ion ($\mu$) obtained from neutron diffraction analysis;\cite{Espeso} 
    Kondo temperature ($T_{K}$) estimated from (i) the magnetic entropy ($S_{mag}$), 
    (ii) the magnetic susceptibility ($\theta_{P}$/10),\cite{Soldevilla98} 
    (iii) quasielastic neutron scattering (QENS),\cite{Espeso} (iv) the minimum 
    in the electrical resistivity ($T_{\rho_{min}}$/5).\cite{Soldevilla98}}
    \begin{ruledtabular}
	\begin{tabular}{cccccccccc}
	    Compound & T$_{C}$ & S$_{mag}$(T$_{C}$) & 
	    T$_{K}$ & T$_{K}$ & T$_{K}$ & T$_{K}$ & $\gamma$ & $\mu$ & CEF\\
	      & (K) & (Rln2) & S$_{mag}$ & QENS & $\theta_{P}$/10 & 
	      T$_{\rho_{min}}$/5 & (mJ/K$^{2}$ mol) & $\mu_{B}$ & $\Delta_{1}$, $\Delta_{2}$ (K)\\
	    \hline
	    CeNi$_{0.1}$Cu$_{0.9}$ & 2.9 & 0.67 & 4.3 & 0.7 & 0.4 & & 131 & 1 & 44, 105\\
	    CeNi$_{0.2}$Cu$_{0.8}$ & 3.0 & 0.58 & 4.8 & 1 & 0.1 & 1.9 & 117 &  & 47, 128\\
	    CeNi$_{0.3}$Cu$_{0.7}$ & 3.1 & 0.45 & 6.1 &  & 0.8 & 2.4 & 114 &  & 50, 116\\
	    CeNi$_{0.4}$Cu$_{0.6}$ & 2.7 & 0.46 & 6.0 & 2 & 1 & 2.8 & 112 & 0.6 & 49, 120\\
	    CeNi$_{0.5}$Cu$_{0.5}$ & 2.5 & 0.58 & 4.6 & & 1.4 & 3.7 & 151 & 0.51 & 55, 146\\
	    CeNi$_{0.6}$Cu$_{0.4}$ & 2.6 & 0.60 & 4.7 & & 1.7 & 4.7 & 154 &  & 62, 135\\
	    CeNi$_{0.7}$Cu$_{0.3}$ & 2.2 & 0.48 & 4.8 & & & & 176 &  & 50, 116\\
	    CeNi$_{0.8}$Cu$_{0.2}$ & 1.8 & 0.19 & 6.6 & & 4.1 & 7.2 & 206 &  & 51, 115\\
	    CeNi$_{0.85}$Cu$_{0.15}$ &  &  & 8.2 & & 6 & 4.5 & 217 &  & 46, 100\\
	    CeNi$_{0.9}$Cu$_{0.1}$ &  &  & 13.8 & & & & 150 &  & 67, 159\\
	    CeNi &  &  & 140 (Ref.~\onlinecite{Gignoux83}) & & & & 65 
	    (Ref.~\onlinecite{Gignoux83}) &  & \\
	\end{tabular}
    \end{ruledtabular}
\end{table*}

Returning to the $\gamma$ evolution for the CeNi$_{1-x}$Cu$_{x}$ 
system displayed in the same plot, we became aware of the coincidence 
of the $\gamma$ variation between both series for Ni content 
larger than 0.5, whereas the compounds closer to CeCu follow a 
different trend from the CeNi$_{y}$Pt$_{1-y}$ series.
Thus, that comparison allows dividing the behavior of the series 
into two main composition regimes. 
Hence, the first regime would range from CeCu to CeNi$_{0.4}$Cu$_{0.6}$, 
where the values of $\gamma$ do not significantly change. This region 
corresponds to the composition regime where the evolution from 
an AFM (CeCu) to FM (CeNi$_{0.4}$Cu$_{0.6}$) ground state occurs, giving 
rise to complex ordered spin structures as has been established 
from neutron diffraction.\cite{Espeso} This evolution is considered 
to be caused by the enhancement of the ferromagnetic interactions 
closely related to the change in the density of states at the 
Fermi surface driven by the differences in d-electron density 
between Cu and Ni\cite{Hernando} and it has been also found 
in similar RNi$_{1-x}$Cu$_{x}$ series with 
light\cite{Soldevilla04} and heavy Rare-Earths.\cite{Gignoux76}

The evolution from CeNi$_{0.4}$Cu$_{0.6}$ to CeNi defines the second 
regime. Once the FM ground state sets in, the competition between 
RKKY, predominantly ferromagnetic, and Kondo interactions leads 
the system from a magnetic localized to a delocalized Pauli Paramagnetic 
state (CeNi), and this competition defines the second regime. 
In that regime, the coefficient $\gamma$ is enhanced for compounds 
with increasing hybridization favored by the decreasing cell 
volume, reaching values of more than 200 mJ/K$^{2}$ mol around the 
$x$=0.15 composition. 

Let us come back to the cell volume 
variations of both series displayed in Fig.~\ref{fig1}. In spite of 
the change in crystallographic structure existing in CeNi$_{1-x}$Cu$_{x}$, 
the cell volume evolves similarly for both series, having both CePt and 
CeCu nearly the same value of the cell volume. They do, 
however, have to be distinguished.
While in CeNi$_{1-x}$Cu$_{x}$ the random substitution on the 
nonmagnetic site Ni/Cu not only changes the cell volume but also 
modifies the electronic state (by changing the number of conduction 
electrons), in the CeNi$_{y}$Pt$_{1-y}$ case the Ni/Pt substitution 
only modifies the cell volume. However, the evolution of $\gamma$ 
is the same in both series in the thus defined second regime in 
CeNi$_{1-x}$Cu$_{x}$. The evolution of $\gamma$ was explained in CeNi$_{y}$Pt$_{1-y}$ 
as driven by the competition between RKKY and Kondo interactions 
under an internal molecular field\cite{Blanco94} according 
to;
$$\gamma \propto \frac{T_{K}}{T_{K}^{2}+H^{2}}$$
where $H$ is the Zeeman energy of the magnetic moment associated 
with a Ce$^{3+}$ ion within the molecular field created by the other 
ions.

Then, it is deduced from the previous comparison that the 
electronic effects do not play a predominant role in the balance 
between RKKY and 4f-conduction band hybridization of the second 
regime. It can thus be concluded that the driving parameters 
of the $\gamma$ behavior in this region are the interatomic 
distances rather than the electronic effects.

\subsection{Estimate of $C_{mag}$}
The magnetic contribution to the specific heat C$_{mag}$ has been 
estimated for all the compounds by taking as the lattice contribution 
of each magnetic compound that of the isomorphous nonmagnetic 
one after taking into account the mass corrections.\cite{Bouvier} Thus, 
the specific heat of LaNi\cite{Blanco} has been taken as 
the nonmagnetic reference for the CrB-type structure compounds 
($x$=0.1 and 0.15) while for the rest of the compositions (FeB-type 
structure) the specific heat of YNi was used. Using this procedure, 
$C_{mag}$ reflects any effect related to the magnetic order transition 
and/or the temperature dependence of the electronic term.

\subsubsection{$C_{mag}$ in the low temperature regime}
\begin{figure}
    \centerline{\includegraphics[width=7cm]{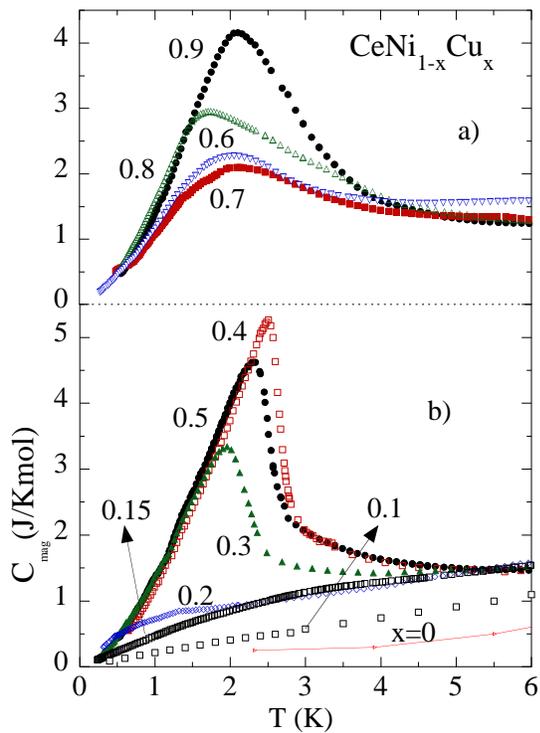}}
    \caption{\label{fig5} $C_{mag}$ versus temperature in the low 
    temperature regime plotted separately for the two composition 
    regions defined: a) 0.9$\leq$$x$$\leq$0.6 or region where the system 
    evolves from AFM to FM ground state, b) $x$$\leq$0.6 or region 
    where hybridization effects and disorder are predominant 
    (for details see text).}
\end{figure}

Fig.~\ref{fig5} displays $C_{mag}$ as a function of the temperature in 
the low temperature range for both regimes as defined in the 
previous section.

Thus, Fig.~\ref{fig5}a illustrates the region 1$>$$x$$>$0.6 where 
the evolution from an AFM to a FM ground state occurs due to 
the modifications of the positive and negative RKKY interactions. 
Fig.~\ref{fig5}b shows $C_{mag}$ for compounds with $x$$<$0.6, where 
the magnetic ground state evolves from Ferromagnetism to a Pauli 
Paramagnetic state (CeNi).

 Starting from the Cu rich side, CeNi$_{0.1}$Cu$_{0.9}$ shows an 
anomaly at the critical temperature ($T_{C}$) corresponding 
to the antiferromagnetic transition ($T_{N}$=2.5 K) defined as 
the inflection point above the maximum of the $C_{mag}$ curve. The 
anomaly becomes broader and its maximum value decreases as Ni 
content increases, it being significantly reduced for the $x$=0.6 
compound, which shows a broad hump centered at about 2 K. It is worth 
noting the smooth variation of $C_{mag}$ above the ordering temperature 
for these compounds. These effects are related to the existence 
of a short-range magnetic order well above $T_{N}$. In fact, the competition 
between positive and negative interactions leads to complex ordered 
spin structures, as was corroborated by neutron diffraction.\cite{Espeso} 
In addition, Quasielastic Neutron Scattering 
(QENS) carried out in $x$=0.9, 0.8, and 0.6\cite{Espeso} indicated 
a supplementary inelastic Gaussian contribution persisting up 
to $\sim$3$T_{C}$, which becomes more important as Cu content 
increases (i. e., as the exchange interactions became more significant). 
All these features support the predominant role played by RKKY 
interactions in setting the magnetic ground state in this regime.

We note a perceptible change in $C_{mag}$ between $x$=0.6 and $x$=0.5. 
Instead of a hump, CeNi$_{0.5}$Cu$_{0.5}$ exhibits a clear $\lambda$-type 
anomaly, with a critical temperature $T_{C}$=2.3 K.
This anomaly becomes even sharper for CeNi$_{0.6}$Cu$_{0.4}$ and 
shifts slightly up to T$_{C}$=2.5 K. For 
the following compounds, the anomaly broadens and its maximum 
value decreases with the increasing Ni content. Thus, for compositions 
with $x$$<$0.2, the anomaly in $C_{mag}$ cannot be distinctly defined.

The temperature corresponding to the maximum of the magnetic 
specific heat coincides with the freezing temperature ($T_{f}$) 
determined by AC magnetic susceptibility measurements, which 
for the compounds in this second regime was associated with a 
``spin-glass-like'' state developing at temperatures above the 
long-range ferromagnetic order.\cite{Soldevilla,GomezSal01} Microscopic 
studies, performed with $\mu$SR in order to shed light on the 
nature and origin of these transitions, claimed for an intermediate 
inhomogeneous magnetic state above T$_{f}$ evolving into a long-range 
ordered state at lower temperatures. However, from the specific 
heat point of view, no other transitions were observed at temperatures 
below the maxima. This feature agrees with the idea of a continuous 
evolution (better than a second order phase transition) to the 
long-range ordered state found from neutron and muon results at 
low temperatures.\cite{Espeso,Marcano} The first 
calorimetric studies presented in 
CeNi$_{0.4}$Cu$_{0.6}$ revealed a very narrow peak at 1 K.\cite{GomezSal}
The actual measurement in this composition is identical to the
already published one except for that feature. We should bear in mind
that Ref.~\onlinecite{GomezSal} was the first in-depth study on one
composition of the CeNi$_{1-x}$Cu$_{x}$ series and, at that point
when even the magnetic phase diagram of the series was unknown,
we were not aware of the enormous relevance of a meticulous 
characterization of the samples. In fact, the sample reported in
Ref.~\onlinecite{GomezSal} was melted in an induction furnace and 
its degree of inhomogeneity was not as well characterized as in
the present study.

\begin{figure}
    \centerline{\includegraphics[width=7cm]{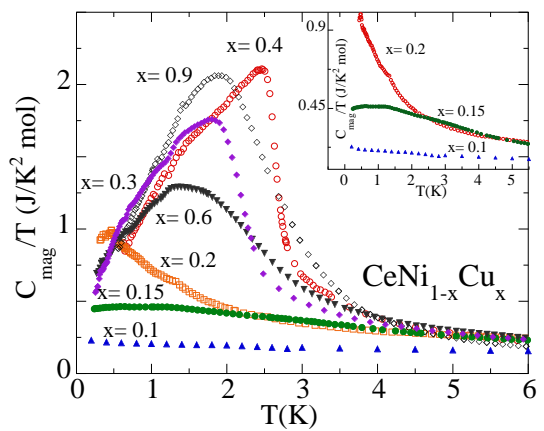}}
    \caption{\label{fig6} $C_{mag}/T$ vs. $T$ representation of the 
    specific heat for CeNi$_{1-x}$Cu$_{x}$ in the compositions 
    studied. The inset shows the variation for the compounds near 
    the magnetic-nonmagnetic crossover $x$=0.1, 0.15, and 0.2.}
\end{figure}

Taking a look at the very low temperature side of the $C_{mag}/T$ 
vs $T$ curves (Fig.~\ref{fig6}), we can observe an almost constant behavior 
for CeNi$_{0.85}$Cu$_{0.15}$ and CeNi$_{0.9}$Cu$_{0.1}$. This fact indicates 
the Fermi-liquid character of these compounds. It should be 
remarked that the Fermi-liquid theory should only be applied 
at very low temperatures where it describes the low lying excitations 
of the quasiparticles (interacting fermions). However, it is 
usually assumed that this label is extended to compounds presenting 
a constant electronic specific heat and magnetic susceptibility 
coefficient and a $T^{2}$ dependence of the electrical resistivity. 
This is the case of the $C_{mag}/T$ corresponding to the $x$=0.1 
and 0.15 compounds.

In f-electron compounds, strong electronic correlations can lead 
to anomalous low-temperature properties indicating that the system 
changes from a nonmagnetic to a magnetic ground state as a certain 
parameter such as concentration or pressure is tuned. Typically, 
the NFL behavior is associated to a diverging 
specific heat $C_{mag}/T$ when approaching $T$=0 K, while the
Fermi-liquid theory predicts a constant value.\cite{Coleman}

In CeNi$_{1-x}$Cu$_{x}$, the whole set of measurements traces a crossover 
from magnetic order to Pauli paramagnetic behavior, thereby crossing 
some critical concentration range where long-range order is expected 
to vanish. Therefore, one might ask if NFL behavior 
might arise in this system. In that sense, what is remarkable is the 
divergence of the $C_{mag}/T$ curve at low temperatures for the 
$x$=0.2 compound.

We are tempted to attribute the origin of this divergence to its being 
due to the critical concentration where long-range magnetic order 
is suppressed, and then presenting a NFL behavior 
as in other series exhibiting, in principle, a similar scenario. 
However, the existence of a transition at lower temperature 
cannot fully be ruled out. The low temperature dependence of $C_{mag}/T$ 
must be carefully analyzed in this case. In fact, as we mentioned 
in the introduction, very low temperature studies by both macroscopic 
(ac and dc measurements)\cite{Marcano03} and microscopic ($\mu$SR) 
techniques\cite{Marcano}
in the compositions around $x$=0.2 revealed 
that the magnetic moments are not yet exhausted. However, long-range 
magnetic order, if it exists, cannot be detected by neutron 
diffraction (within the resolution of the available experiments) 
since we are dealing with very reduced values of the magnetic 
moment. Both experiments point towards some kind of short-range 
ordered states and reveal evidence of spin-glass arrangements 
even for the $x$=0.1 compound.\cite{Marcano04} Then, on the basis of 
all these results we have to discard an intrinsic electronic 
effect leading to NFL or Quantum Phase transition 
around the $x$=0.2 compound. The divergence on $C_{mag}/T$ could 
be related to the existence of a small anomaly at lower temperatures 
as was found at very low temperature in AC susceptibility 
measurements.\cite{Marcano03}

\subsubsection{Low temperature magnetic entropy}
\begin{figure}
    \centerline{\includegraphics[width=7cm]{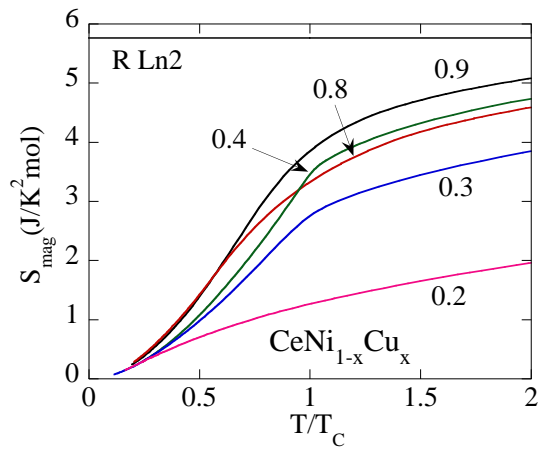}}
    \caption{\label{fig7} Magnetic entropy $S_{mag}$ for the 
    CeNi$_{1-x}$Cu$_{x}$ versus reduced temperature ($T$/$T_{C}$) 
    for some selected compounds for clarity.}
\end{figure}

The magnetic entropies have been calculated from
$$S_{mag}=\int_0^T {{{C_{mag}} \over T}dT}$$
and are shown in Fig.~\ref{fig7} versus the reduced temperature 
($T$/$T_{C}$) for the most significant compounds.

The first point to be discussed in the entropy analysis is the 
shape of these curves below $T_{C}$. As has been pointed out 
in other series such as RGa$_{2}$ (Ref.~\onlinecite{Blanco92}) and 
RNi$_{2}$Si$_{2}$ (Ref.~\onlinecite{Barandiaran}), 
this shape is closely related to the magnetic structure and the 
thermal demagnetization processes. The more simple the structure 
is with well-defined anisotropy, the sharper the increase of 
the entropy is, defining a clear change of 
slope at $T_{C}$. This is the case shown in Fig.~\ref{fig7} 
where for the AFM complex structures ($x$=0.8 and
0.9) a slow and progressive increase is observed, whereas for 
the FM one ($x$=0.4) the increase is sharper and a net knee is 
observed at T$_{C}$. It must be stressed, however, that in 
this CeNi$_{1-x}$Cu$_{x}$ 
series these variations are quite smooth even in the case of 
FM compounds, indicating a high degree of magnetic disorder.

\begin{figure}
    \centerline{\includegraphics[width=7cm]{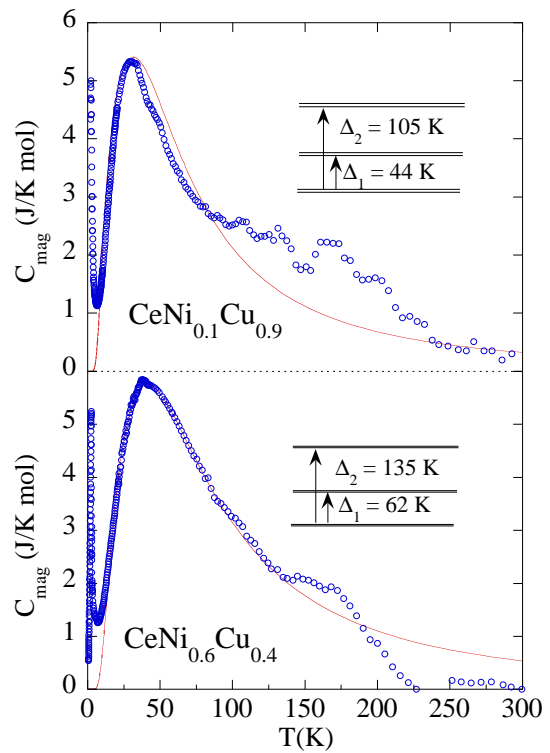}}
    \caption{\label{fig8} Magnetic contribution to the specific heat 
    as a function of temperature for CeNi$_{0.1}$Cu$_{0.9}$ and 
    CeNi$_{0.6}$Cu$_{0.4}$. Solid lines correspond to the 
    paramagnetic CEF contribution (see text). The arrows indicate 
    the main transitions in the CEF level scheme.}
\end{figure}

In all the compounds the magnetic entropy at $T_{C}$ is lower than 
R$\cdot$ln2=5.76 J/K mol , the value expected from a doublet ground state, 
which is the one presented by the Ce$^{3+}$ ion under an orthorhombic 
framework. The reasons for such a reduction are at least two: 
The first one is the complexity of the magnetic order with the 
existence of short-range interactions above $T_{C}$; this should 
be the main effect in compounds close to the CeCu limit. The 
value of the entropy at $T_{C}$ decreases as the structure become 
more complex (RKKY competing interactions being the predominant 
effect). The second is the Kondo effect, which reduces the value 
of the magnetic moment that is the dominant effect once the ferromagnetic 
order is established ($x$$<$0.6) and the Kondo interactions 
become more important. This reduction of the magnetic moment 
has been confirmed by neutron and muon experiments.\cite{GomezSal,Marcano}

From this last assumption a $T_{K}$ value can be estimated considering 
the reduction of the magnetic entropy at $T_{C}$ from the value 
of R$\cdot$ln2.\cite{Yashima} Using a simple two-level model with 
an energy splitting of $k_{B}T_{K}$, we can calculate 
the reduced entropy at $T_{C}$ as:
$${{\Delta S} \over R}=\ln \left[ {1+\exp \left( {{{-T_K} \over {T_C}}} \right)} \right]+
{{T_K} \over {T_C}}\left[ {{{\exp \left( {{{-T_K} \over {T_C}}} \right)} \over 
{1+\exp \left( {{{-T_K} \over {T_C}}} \right)}}} \right]$$

This expression must be applied for transitions at $T_{C}$ to 
long-range magnetic order which is not really the case in our compounds 
with $x$$\leq$0.6. In this case the specific heat anomaly at 
$T_{C}$ is related to the spin freezing temperature. However, the 
frozen state in these samples is a cluster-glass with
ferromagnetic correlations inside the clusters. So, even if some
fraction of the entropy is invested in the disorder among the 
clusters, this approach can give us an estimate of the Kondo 
temperatures in order to follow their compositional evolution.

For the compounds with no anomaly in $C_{mag}$ ($x$$<$0.2), the 
estimation of $T_{K}$ requires another kind of analysis. 
From the Sacramento and Schlottmann calculations for 
a J=1/2 impurity,\cite{Sacramento} one could estimate $T_{K}$ as 
the value at which 45\% of the total R$\cdot$ln2 entropy is recovered. 

Taking into account the perturbation induced by the short-range 
order correlations and the two different methods used to estimate 
the $T_{K}$ values, these results must be looked at with care.

Table~\ref{table2} summarizes the values obtained for $T_{K}$ in 
each composition. 
In the same table we also present the values of $T_{K}$ estimated 
from susceptibility\cite{Soldevilla98} as 
${\left|{\theta _P}\right|}/10$, QENS\cite{Espeso}, 
and resistivity as 
T$_{\rho_{min}}$/5.\cite{Soldevilla98}

It is clear that the right estimate of $T_{K}$ comes from QENS\cite{Espeso} 
where the quasielastic signal is directly related to the Kondo 
temperature. The values of $T_{K}$ obtained from the magnetic entropy 
are clearly overestimated due to the entropy associated to the 
cluster-glass formation. 

The most interesting feature in all these estimates of $T_{K}$ is 
that, in spite of the different magnitudes due to the different 
time scales of the techniques,\cite{Blanco94,Galera,Gignoux91} the 
same relative variation of $T_{K}$ is found with increasing Ni content.

\subsubsection{$C_{mag}$ above $T_{C}$}
Apart from short-range order correlations that might be present 
above the critical temperature, the contribution to $C_{mag}$ above 
$T_{C}$ arises from CEF. Due to the low symmetry site occupied by 
the Ce$^{3+}$ ions in this system, the J=5/2 multiplet splits into 
three doublets, with the excited states separated from the ground 
state by energy gaps $\Delta_{1}$ and $\Delta_{2}$. A Schottky-type 
anomaly is then expected. The hump around 50K appearing in all 
the studied compounds of this series has been analyzed considering 
this CEF scheme. The values $\Delta_{1}$ and $\Delta_{2}$ that 
correspond to the best Schottky-type fit to $C_{mag}$ are listed 
in Table~\ref{table2}. 

The values obtained for the energy gaps are similar for all 
the compositions with FeB type of structure. Moreover, the level 
scheme found for the CrB compositions ($x$=0.15 and 0.1) is close 
to that corresponding to the FeB-type ones. This result indicates 
that the Ni/Cu dilution does not significantly modify the CEF 
splitting over the series.  

Special attention must be paid to the CrB compositions. On the 
one hand, the presence of Schottky-type anomalies supports the
fact that the magnetic moment in Ce$^{3+}$ remains localized
still for those concentrations close to CeNi. On the other hand,
the values found 
for the samples with CrB and FeB-type structures reveal that the 
change in the structure does not introduce significant changes 
in the CEF sensed by the Ce$^{3+}$ ions. This fact indicates the 
similarity between both structures as was already pointed 
out in Sect.~\ref{Samples}.

In contrast, it is worth noting that for the intermediate valence 
compound CeNi, a broad contribution centered around 140 K was 
observed in previous works \cite{Gignoux83}, it being related to 
spin fluctuations present in compounds close to the onset of 
ferromagnetism.

The magnetic contribution to the specific heat in the paramagnetic 
range together with the best fit obtained for the Schottky anomaly 
and the associated level scheme are illustrated in Fig.~\ref{fig8} as 
an example for $x$=0.9 and $x$=0.4.

\begin{figure}
    \centerline{\includegraphics[width=7cm]{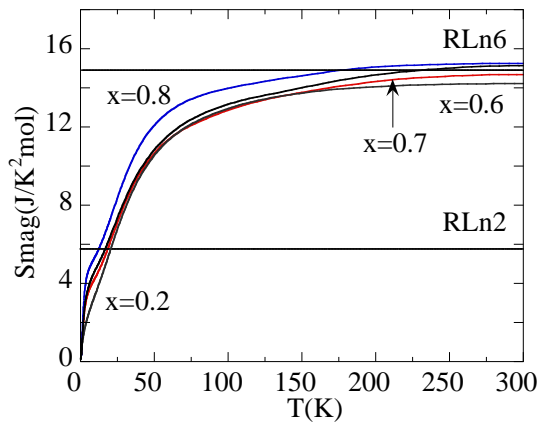}}
    \caption{\label{fig9} Magnetic entropy $S_{mag}$ versus 
    temperature extended to all temperature range studied 
    for CeNi$_{1-x}$Cu$_{x}$.}
\end{figure}

Finally, the magnetic entropy $S_{mag}$ calculated from the experimental 
$C_{mag}$ variation for the 0.2 K$<$$T$$<$300 K temperature range 
is shown in Fig.~\ref{fig9} for some selected compounds. For increasing 
temperature, $S_{mag}$ progressively increases due to the thermal 
population of the excited CEF levels, reaching saturation 
around 250 K. The saturation values are close to the theoretical 
value R$\cdot$ln6=14.89 J/K$^{2}$ mol, as corresponds 
to the magnetic entropy of a Ce$^{3+}$ 
ion with a total angular momentum J=5/2.

\section{Discussion}
The actual calorimetric study has provided useful information 
about the complex magnetic ground state found in this system. 
This section is devoted to analyzing the behavior of the series, 
based on both the present and the previous studies carried out 
so far.

First, we must remember the division of the series into two regimes 
with respect to the heat capacity findings. Although the competing 
interactions are present throughout the series, heat capacity 
reflects that the interplay of the interactions and its magnitude 
are tuned by the composition. Thus, we have defined a first regime 
where the competing exchange interactions are 
predominant (0.9$\leq$$x$$<$0.6),
giving rise to a change from antiferromagnetism to ferromagnetism. 
The second regime is characterized by the increasing Kondo effect 
driven by the decreasing cell volume. As the Kondo effect 
becomes more important, the Ce magnetic moments become smaller 
and the effective magnetic interaction decreases.

In this second regime the general behavior is the same as that 
expected for a Kondo lattice system and the electronic 
coefficient, $\gamma$, evolves similarly as in the 
CeNi$_{y}$Pt$_{1-y}$ series,\cite{Blanco94} a clear 
example of a Doniach predicted behavior.

The maximum value of $\gamma$ corresponds to the composition 
where Kondo interactions overcome the molecular field effects 
and then, long-range magnetic order is hard to reach. For 
compounds presenting higher Ni content, $\gamma$ behaves as 
in the nonordered heavy fermion compounds
$\gamma\propto\frac{1}{T_{K}}$. In our case, however, important 
magnetic disorder effects still 
prevail. Furthermore, the magnetic moments are not completely 
exhausted and they can arrange at very low temperatures.

\begin{figure*}
    \centerline{\includegraphics[width=15cm]{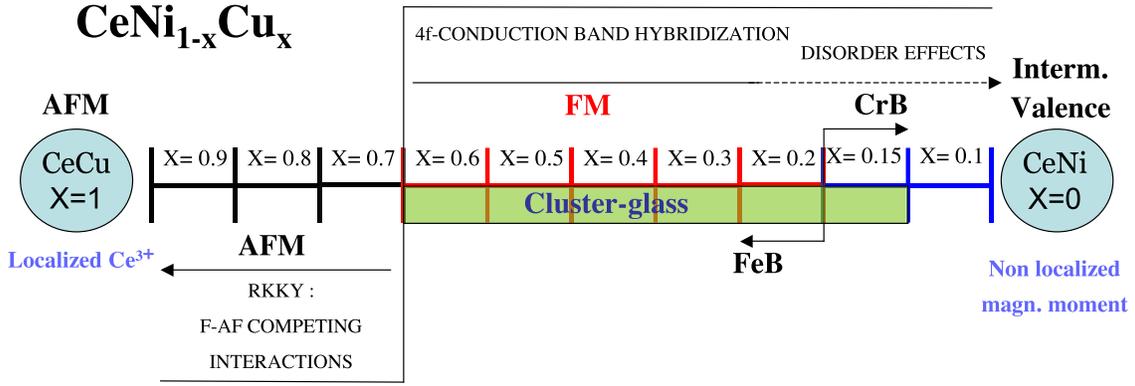}}
    \caption{\label{fig10} Schematic evolution of both the magnetic 
    behavior and structural characteristics of CeNi$_{1-x}$Cu$_{x}$. 
    The two different regimes deduced from these calorimetric 
    studies are indicated together with the predominant interaction.}
\end{figure*}

The complex magnetic behavior and the evolution of the interactions 
in this series can be condensed in the sketch picture displayed 
in Fig.~\ref{fig10}.

Previous bulk data suggested the existence of a spin-glass-like 
phase (cluster-glass in Fig.~\ref{fig10}) appearing in 
a certain compositional range of this series at the freezing 
temperature $T_{f}$.\cite{Soldevilla,GomezSal01} In addition, long-range 
ferromagnetic order at very low temperatures has been confirmed 
by neutron diffraction and $\mu$SR studies at least down to 
the composition with $x$=0.2.\cite{Marcano,Thesis} The low temperature 
specific heat for these compounds exhibits marked anomalies at 
the freezing temperature $T_{f}$ detected by ac susceptibility. 
The shape of these anomalies, however, differs from what is expected 
around the freezing temperature in a canonical spin-glass.

So far, a lot of experimental data on the magnetic heat capacity 
behavior of ``spin-glass'' systems are available and few analytical 
expressions have been used to reproduce the data for the different 
samples.\cite{Binder,Mydosh} Furthermore, the ``spin-glass'' 
label has been used to refer to many different 
situations ranging from canonical 
spin-glass (i.e. individual spins) to magnetic ground states 
close to superparamagnetism (i.e. noninteracting magnetic spin 
clusters). That is the reason why some of the spin-glass properties 
do not seem to be universal. In particular, it is usually assumed 
that the magnetic contribution to the specific heat shows no 
sharp anomaly at the freezing temperature, as found in other heavy 
fermion intermetallic compounds with spin-glass-like behavior: 
U$_{2}$PdSi$_{3}$ (Ref.~\onlinecite{Li}), 
URh$_{2}$Ge$_{2}$ (Ref.~\onlinecite{Sullow}), and more 
recently CeNi$_{2}$Sn$_{2}$ (Ref.~\onlinecite{Tien}). 
Our data do not conform 
with this common phenomenology although the magnetic macroscopic 
measurements clearly point to a spin-glass-like behavior.

\begin{figure}
    \centerline{\includegraphics[width=7cm]{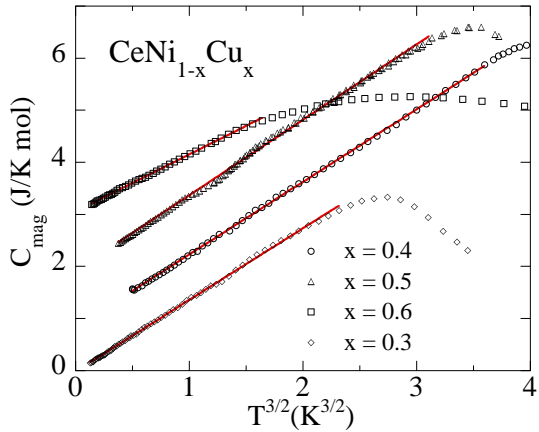}}
    \caption{\label{fig11} Temperature dependence of the magnetic 
    contribution to the specific heat $C_{mag}$ for the FM 
    compounds $x$=0.3, 0.4, 0.5, and 0.6 in the form $C_{mag}$ 
    versus $T^{3/2}$. The data corresponding to different 
    compounds are shifted for clarity.}
\end{figure}

Furthermore, the magnetic entropy reached at the spin freezing 
temperature is estimated to be 45-60\% of the theoretical value 
Rln2 corresponding to the full degeneracy of the magnetic ground 
state (see Table~\ref{table2}). That percentage is slightly higher than 
those reported for the crystalline dilute spin-glass systems 
MnCu, or AuFe (22-33\%),\cite{Wenger} even taking into account
the entropy reduction due to the Kondo effect in our samples. 
Compared with our data, 
a similar specific heat and magnetic entropy behavior was also 
found in amorphous Gd$_{33}$Al$_{67}$ (Ref.~\onlinecite{Coey}) 
and Er$_{50}$Ni$_{50}$ (Ref.~\onlinecite{Hattori})
alloys. In addition, for those compounds the 
magnetic specific heat showed a $T^{3/2}$ temperature dependence, 
indicating a collective excitation. These results indicate that 
ferromagnetic-like spin waves can be excited in these amorphous 
alloys, although the overall magnetic structure is not ferromagnetic 
but rather spin-glass. Fig.~\ref{fig11} displays the temperature dependence 
of our magnetic specific heat in the form $C_{mag}$ versus $T^{3/2}$ dependence 
for the ferromagnetic compounds $x$=0.3, 0.4, 0.5, and 0.6. It is 
clear that $C_{mag}$ follows the $T^{3/2}$ dependence up to about 0.7$T_{f}$ 
in a similar manner as observed in the amorphous compounds cited 
above. The existence of these ferromagnetic correlations yields 
a faster enhancement of the magnetic entropy than that of a canonical 
spin-glass (linear with temperature). Then, the maximum value 
of the specific heat has to be reached at a lower temperature 
than the 1.6$T_{f}$ expected value,\cite{Binder,Mydosh} as happens 
in our series.

Neutron diffraction experiments at very low temperature have 
been conclusive to ascertain the existence of long-range ferromagnetic 
order in $x$=0.6, 0.5, and 0.4.\cite{Espeso,Thesis} Neutron scattering 
is primarily a probe of long-range correlations. On the other 
hand, $\mu$SR spectroscopy is a much more local probe than 
neutron scattering, and needs no large coherence length. Consequently, 
it is particularly sensitive to short-range order and other forms 
of disordered magnetism.\cite{Kalvius} Our $\mu$SR results 
indicate the existence of an inhomogeneous magnetic state at 
temperatures above the freezing temperature T$_{f}$ 
(seen by ac-susceptibility) that runs into the paramagnetic 
regime. The basic feature of this intermediate 
state is the coexistence of ordered and nonordered fractions. 
The latter decreases as the temperature is lowered until T$_{f}$ 
is reached. Below T$_{f}$, long-range order, from the $\mu$SR
perspective,\cite{Kalvius} prevails. Nevertheless, 
it must be stressed that this long-range ordered state detected 
by muons presents strong local magnetic disorder as reflected 
by the wide field distribution on the muon site arising from 
those analyses. This distribution leads to the fast damping of 
the oscillatory pattern observed in the series.\cite{Marcano} 

It is worth remembering the difference in the time window for the 
study of spin dynamical processes when comparing the different 
experimental techniques used so far (i.e. ac susceptibility, 
specific heat, neutron diffraction, and $\mu$SR spectroscopy). 
Each technique can access a different characteristic range 
of the rate of spin fluctuations and hence provides different 
information concerning dynamical processes developing in the 
system. The combination of all the techniques, however, yields 
a complete dynamical picture of the system since $\mu$SR bridges 
the gap between neutron scattering on the one side and bulk magnetic 
measurements on the other.\cite{Kalvius}

Considering our experimental results as a whole, we can propose 
a convenient description of our system. Although x-ray, neutron 
diffraction, and scanning microscopy reflect a clear homogeneity 
of the samples at the micrometric level, we can ensure that inhomogeneities 
in the random distribution of Cu and Ni appear at the nanometric 
scale. When temperature is lowered from the paramagnetic state, 
zones or clusters of magnetic moments develop due to the increasing 
short-range magnetic interactions. These clusters freeze at $T_{f}$ 
losing their dynamical aspect. At lower temperatures the magnetic 
clusters interact among one another and percolate giving rise 
to a long-range ferromagnetic state at very low temperatures, 
in agreement with the ferromagnetic excitations detected by specific 
heat below $T_{f}$. This percolative process scenario, reminiscent 
of that proposed by Dagotto in manganites,\cite{Dagotto} is 
also present in many substitutional systems such as semiconductors, 
high $T_{C}$ superconductors, etc.\cite{Mayr} The CeNi$_{1-x}$Cu$_{x}$ 
is, thus, a clear example of Strongly Correlated Electron System 
presenting a percolative scenario.

\begin{figure}
    \centerline{\includegraphics[width=7cm]{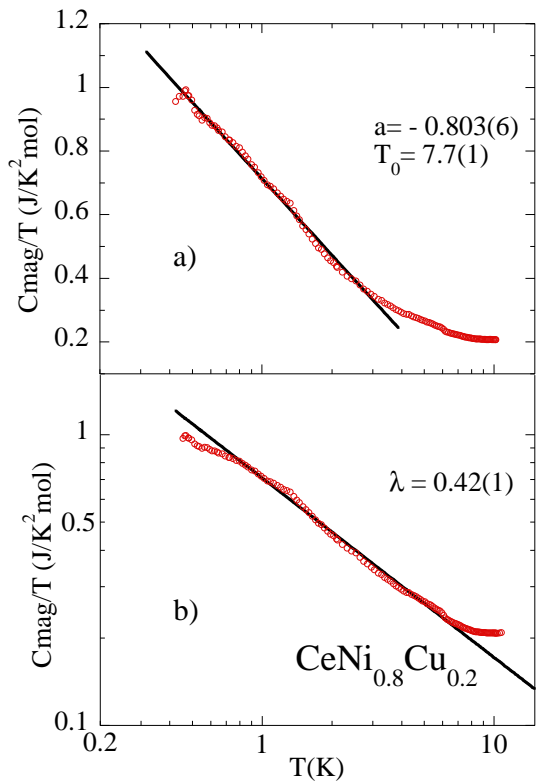}}
    \caption{\label{fig12} a) $C_{mag}/T$ vs log$T$ and b) 
    log($C_{mag}/T$) vs log$T$ plots for the $x$=0.2 compound. 
    The reported $a$, $T_{0}$, and $\lambda$ values are the 
    obtained parameters from the 
    a) $C_{mag}/T \propto a $log ($T/T_{0}$) law and 
    b) $C_{mag}/T \propto T^{-1+\lambda}$ law.}
\end{figure}

In the compound with $x$=0.2, the inhomogeneous magnetic state, 
where ordered and nonordered fractions coexist, extends over a 
wider temperature range. Full long-range ordered state has been 
only detected by $\mu$SR and ac susceptibility below 
T$\sim$0.5 K.\cite{Marcano,Marcano03} At this point we must 
remember that this compound was 
reported to be on the crossover of the localized-nonlocalized 
magnetism.\cite{Soldevilla} In fact, the specific heat measurements 
do not show any sharp transition as in the other compounds. Furthermore, 
recent neutron diffraction studies do not detect any magnetic 
contribution within the experimental limits.\cite{Thesis} 
These results evidence how reduced the Ce magnetic moments are 
in this alloy. Being so close to the disappearance of the magnetic 
moment and regarding the $C_{mag}/T$ behavior presented in the 
previous section, one could be tempted to look for a NFL 
behavior defining a QCP in this composition. In that sense, 
Fig.~\ref{fig12} shows the corresponding fits of $C_{mag}/T$ for this compound 
to a log$T$ law, as observed for the case of 
magnetic fluctuations\cite{Moriya,Millis,Ishigaki} and a 
$T^{-1+\lambda}$ law as proposed by Castro Neto,\cite{CastroNeto} 
who considered the existence of Griffiths phases consisting of 
magnetic clusters embedded in a nonmagnetic matrix. 

Although the agreement factor of these fits is comparable to 
other systems reported as non-Fermi-liquids,\cite{Maple} in 
the present case we have powerful reasons to discard this option. 
On the one hand, ac susceptibility shows the presence of a transition 
at very low temperatures as we pointed out above\cite{Marcano03} 
and, on the other hand, the hysteresis loops in $x$=0.15 and 0.1 
compounds observed at very low temperatures\cite{Marcano04} suggest 
the existence of a ``cluster-glass'' state even for those compounds 
that macroscopically behave as Fermi-liquids with almost constant 
$C_{mag}/T$ value. 

In our opinion the scenario for this series has to be understood 
without considering a QCP situation, but rather as only a consequence 
of the increasing hybridization in an intrinsic inhomogeneous 
state that leads to a cluster-glass situation. The more important 
the reduction of the magnetic moment is, the more difficult the 
stability of a long-range ordered state is. This interpretation 
resembles the situation described by Coleman,\cite{Coleman99,Schroder} 
where the magnetic moments still persist in the NFL 
regime such as is the case in MnSi.\cite{Pfleiderer}

The possibility of the scenario proposed here could be extended 
to other substitutional systems such as UCu$_{5-x}$Pd$_{x}$ 
(Ref.~\onlinecite{Bauer,Booth}) and more recently 
UPd$_{2-x}$Sn$_{x}$ (Ref.~\onlinecite{Maksimov}) and 
CePtSi$_{1-x}$Ge$_{x}$ (Ref.~\onlinecite{Young}) where intrinsic disorder effects 
were detected. This fact points out the importance of careful analysis 
of such systems close to the evanescence of the long-range magnetic 
order (QCP). The quantum effects due to new 
electronic ground states must be confirmed by studies carried 
out in good single crystal samples and avoiding substitutional 
effects.

\section{Conclusions}
CeNi$_{1-x}$Cu$_{x}$ is a complex system that presented many open questions 
concerning its magnetic behavior. The analysis of the specific 
heat has shed light on most of those intriguing experimental 
facts. In this way, it has been possible to distinguish two main 
concentration regimes: the first one close to the Cu-rich side, 
where the exchange interactions are predominant and the magnetic 
behavior resembles our findings in other RNi$_{1-x}$Cu$_{x}$ series 
(i.e. NdNi$_{1-x}$Cu$_{x}$ (Ref.~\onlinecite{Soldevilla04})) 
leading to changes from 
AFM to FM; the second one with ferromagnetic behavior, but governed 
by Kondo interactions tending to reduce the Ce magnetic moment 
that vanishes in the CeNi compound in agreement with the Doniach 
model, in a similar way as already found in 
CeNi$_{y}$Pt$_{1-y}$.\cite{Blanco94,Doniach}
We have also carefully investigated the possible existence of 
a NFL behavior for compositions close to the apparent vanishing 
of the long-range order and its relation to a QCP. Considering 
the analysis of $C_{P}$ alone, we can report an enhanced Fermi-liquid 
($\gamma$=constant with temperature) for compounds with $x$$<$0.2, 
and a divergence in the $C_{P}/T$ vs $T$ plot for $x$=0.2 that could 
be reasonably well reproduced by NFL models. However, supplementary 
information obtained by ac and dc susceptibility measurements confirm 
the existence of hysteresis loops at low temperature indicating 
that the magnetic moments are not fully exhausted in those compounds. 
QCP scenario is then not required to account for that behavior 
in this series.

A special effort has been made to understand the magnetic 
contribution of $C_{P}$ related to the spin-glass state or the short-range 
magnetic ordered state. For this purpose we have considered previous $\mu$SR, 
neutron diffraction, and $\chi_{AC}$ results, which provide different 
time scales of those dynamical processes. The critical 
temperature, $T_{C}$, represents the 
temperature where ferromagnetic correlations are established, although 
full long-range order only occurs at lower temperatures. Considering 
the overall results it has been possible to propose a convenient 
description of our system. Its magnetic behavior can be fully 
described by the existence of interacting magnetic clusters that 
percolate, giving rise to a long-range magnetic state at very 
low temperatures.

Our results allow us to extend the large variety of examples of 
systems presenting intrinsic inhomogeneities that reach the 
long-range magnetic order by percolating processes (semiconductors, 
manganites, High-$T_{C}$ superconductors, etc.) to 
Strongly Correlated Electron metals underlying those intrinsic disorder 
effects, which cannot be avoided and are present in mixed or substitutional 
systems.

\begin{acknowledgments}
This work was supported by the MAT2003-06815, the FERLIN-ESF program, and
the Fundaci\'on Ram\'on Areces. N. Marcano and J. Herrero-Albillos
acknowledge MEC for their FPU Ph.D. grants. 
Fruitful discussions with G. M. Kalvius are also acknowledged. We are
indebted to David Culebras for his invaluable technical assistance.
\end{acknowledgments}



\begin{thebibliography}{}
    \bibitem{Gopal} E. S. R. Gopal, in \textit{Specific heats at 
    low temperatures}, (Heywood books, London, 1966).
    
    \bibitem{Coleman} See for instance: \textit{Proceedings of the 
    Conference on NFL Behaviour in Metals,} edited by P. Coleman, 
    M. B. Maple, and A. J. Millis, Santa Barbara $[$J. Phys: Condens. 
    Matter \textbf{8}, Number 48 (1996)$]$.
    
    \bibitem{Moriya} T. Moriya and T. Takimoto, J. Phys.  Soc.  
    Japan \textbf{64}, 960 (1995).
    
    \bibitem{Coleman99} P. Coleman, Physica B \textbf{259-261}, 353 (1999).
    
    \bibitem{Schroder} A. Schr\"{o}der, G. Aeppli, R. Coldea, M. Adams, 
    O. Stockert, H. von L\"{o}hneysen, E. Bucher, R. Ramazashvili, and  
    P. Coleman, Nature \textbf{407}, 351 (2000).
    
    \bibitem{CastroNeto} A. H. Castro Neto, G. Castilla, and B. A. Jones, 
    Phys. Rev. Lett. \textbf{81}, 3531 (1998).
    
    \bibitem{Bernal} O. O. Bernal, D. E. MacLaughlin, H. G. Lukefahr, and 
    B. Andraka, Phys. Rev. Lett. \textbf{75}, 2023 (1995).
    
    \bibitem{Miranda} E. Miranda, V. Dobrosavljevic, and G. Kotliar, Phys. 
    Rev. Lett. \textbf{78}, 290 (1997).

    \bibitem{Espeso} J. I. Espeso, J. Garc\'{\i}a Soldevilla, J. A. Blanco, 
    J. Rodr\'{\i}guez Fern\'{a}ndez, J. C. G\'{o}mez Sal, and M. T. Fern\'{a}ndez 
    D\'{\i}az, Eur. Phys. J. B \textbf{18}, 625 (2000).
    
    \bibitem{GomezSal} J. C. G\'{o}mez Sal, J. Garc\'{\i}a Soldevilla, 
    J. A. Blanco, J. I. Espeso, J. Rodr\'{\i}guez Fern\'{a}ndez, F. Luis, 
    F. Bartolom\'{e}, and J. Bartolom\'{e}, Phys. Rev. B \textbf{56}, 
    11741 (1997).
    
    \bibitem{Soldevilla} J. Garc\'{\i}a Soldevilla, J. C. G\'{o}mez Sal, J. A. 
    Blanco, J. I. Espeso, and J. Rodr\'{\i}guez Fern\'{a}ndez, Phys. Rev. B 
    \textbf{61}, 6821 (2000).
    
    \bibitem{GomezSal02} J. C. G\'{o}mez Sal, J. I. Espeso, J. Rodr\'{\i}guez 
    Fern\'{a}ndez, N. Marcano, and J. A. Blanco, J. Magn. Magn. Mater 
    \textbf{242-245}, 125 (2002).
    
    \bibitem{Marcano} N. Marcano, G. M. Kalvius, D. R. Noakes, J. C. G\'{o}mez Sal, 
    R. W\"{a}ppling, J. I. Espeso, E. Schreier, A. Kratzer, Ch. Baines, 
    and A. Amato, Physica Scripta \textbf{68}, 298 (2003).
    
    \bibitem{Marcano03} N. Marcano, J. I. Espeso, J. C. G\'{o}mez Sal, L. 
    S\'{a}nchez, G. M. Kalvius, C. Paulsen, and Ch. Sekine, Acta Physica 
    Polonica B \textbf{34}, 1477 (2003).
    
    \bibitem{Marcano04} N. Marcano, D. Paccard, J. I. Espeso, J. Allemand, 
    J. M. Moreau, A. Kurbakov, C. Sekine, C. Paulsen, E. Lhotel, and J. C. 
    G\'{o}mez Sal, J. Magn. Magn. Mater \textbf{272-276}, 468 (2004).
    
    \bibitem{Bartolome} J. Bartolom\'{e} and F. Bartolom\'{e}, Phas. Transit. 
    \textbf{64}, 57 (1997).
    
    \bibitem{Miedema} A. R. Miedema, R. F. Wielinga, and W. J. Huiskamp, Physica 
    \textbf{31}, 1585 (1965).
    
    \bibitem{Algra} H. A. Algra, L. J. de Jongh, W. J. Huiskamp, and R.L. Carlin, 
    Physica B \textbf{92}, 187 (1997).
    
    \bibitem{Soldevilla98} J. Garc\'{\i}a Soldevilla, J. C. G\'{o}mez Sal, 
    J. I. Espeso, J. Rodr\'{\i}guez Fern\'{a}ndez, J. A. Blanco, M. T. Fern\'{a}ndez 
    D\'{\i}az, and H. Buttner, J. Magn. Magn. Mater. \textbf{177-181}, 300 (1998).
    
    \bibitem{Gignoux} D. Gignoux and J. C. G\'{o}mez Sal, Phys. Rev. B 
    \textbf{30}, 3967 (1984).
    
    \bibitem{Blanco} J. A. Blanco, J. C. G\'{o}mez Sal, J. Rodr\'{\i}guez 
    Fern\'{a}ndez, M. Castro, R. Burriel, D. Gignoux, and D. Schmitt, 
    Solid State Commun. \textbf{89}, 389 (1994).
    
    \bibitem{Gignoux83} D. Gignoux, F. Givord, R. Lemaire, and F. Tasset, 
    J. Less-Common Met. \textbf{94}, 165 (1983).
    
    \bibitem{Blanco94} J. A. Blanco, M. de Podesta, J. I. Espeso, J. C. G\'omez Sal, 
    C. Lester, K. A. McEwen, N. Patrikios, and J. Rodr\'{\i}guez Fern\'{a}ndez, 
    Phys. Rev. B \textbf{49}, 15126 (1994).
    
    \bibitem{Doniach} S. Doniach, in \textit{Valence Instabilities and Related 
    Narrow Band Phenomena}, edited by R. D. Parks (Plenum, New York, 1976) p. 169.
    
    \bibitem{Hernando} A. Hernando, J. M. Rojo, J. C. G\'{o}mez Sal, and J. M. Novo, 
    J. Appl. Phys. \textbf{79}, 4815 (1996).
    
    \bibitem{Soldevilla04} J. Garc\'{\i}a Soldevilla, J. A. Blanco, J. 
    Rodr\'{\i}guez Fern\'{a}ndez, J. I. Espeso, J. C. G\'{o}mez Sal, M. T. 
    Fern\'{a}ndez-D\'{\i}az, J. Rodr\'{\i}guez Carvajal, and D. Paccard, 
    Phys. Rev. B \textbf{70}, 224411 (2004).
    
    \bibitem{Gignoux76} D. Gignoux and J. C. G\'{o}mez Sal, J. Magn. Magn. Mater. 
    \textbf{1}, 203 (1976).
    
    \bibitem{Bouvier} M. Bouvier, P. Lethuiller, and D. Schmitt, Phys. Rev. B 
    \textbf{43}, 13137 (1991).
    
    \bibitem{GomezSal01} J. C. G\'{o}mez Sal, J. Rodr\'{\i}guez Fern\'{a}ndez, 
    J. I. Espeso, N. Marcano, and J.A. Blanco, J. Magn. Magn. Mater. 
    \textbf{226-230}, 124 (2001).
    
    \bibitem{Blanco92} J. A. Blanco, D. Gignoux, J. C. G\'{o}mez Sal, J. 
    Rodr\'{\i}guez Fern\'{a}ndez, and D. Schmitt, J. Magn. Magn. Mater. 
    \textbf{104-107}, 1285 (1992).
    
    \bibitem{Barandiaran} J. M. Barandiar\'{a}n, D. Gignoux, D. Schmitt, 
    J. C. G\'{o}mez Sal, and J. Rodr\'{\i}guez Fern\'{a}ndez, J. Magn. Magn. 
    Mater. \textbf{69}, 61 (1987).
    
    \bibitem{Yashima} H. Yashima, H. Mori, N. Sato, and T. Satoh, J. Magn. 
    Magn. Mater. \textbf{31-34}, 411 (1983).
    
    \bibitem{Sacramento} P. D. Sacramento and P. Schlottmann, Phys. Rev. B 
    \textbf{40}, 431 (1989).
    
    \bibitem{Galera} R. M. Galera, A. P. Murani, and J. Pierre, Physica B 
    \textbf{156-157}, 801 (1989).
    
    \bibitem{Gignoux91} D. Gignoux, A. P. Murani, D. Schmitt, and M. Zerguine, 
    J. Phys. (France) \textbf{1}, 281 (1991).
    
    \bibitem{Thesis} N. Marcano, PhD. Thesis. University of Cantabria. 
    December 2004.
    
    \bibitem{Binder} K. Binder and A. P. Young, Rev. Mod. Phys. \textbf{58}, 
    801 (1986).
    
    \bibitem{Mydosh} J. A. Mydosh, in \textit{Spin glasses: An experimental 
    introduction}, (Taylor \& Francis, London, 1993).
    
    \bibitem{Li} D. X. Li, Y. Shiokawa, Y. Homma, A. Uesawa, A. D\"{o}nni, T. 
    Suzuki, Y. Haga, E. Yamamoto, T. Honma, and Y. Onuki, Phys. Rev. B 
    \textbf{57}, 7434 (1998).
    
    \bibitem{Sullow} S. S\"{u}llow, G. J. Nieuwenhuys, A. A. Menovsky, J. A. Mydosh, 
    S. A. M. Mentink, T. E. Mason, and W. J. L. Buyers, Phys. Rev. Lett. \textbf{78}, 
    354 (1997); 
    S. S\"{u}llow, S. A. M. Mentink, T. E. Mason, W. J. L. Buyers, G. J. Nieuwenhuys, 
    A. A. Menovsky, and J. A. Mydosh, Physica B \textbf{230-232}, 105 (1997); 
    S. S\"{u}llow, S. A. M. Mentink, T. E. Mason, R. Feyerherm, G. J. Nieuwenhuys, 
    A. A. Menovsky, and J.A. Mydosh, Phys. Rev. B \textbf{61}, 8878 (2000).
    
    \bibitem{Tien} C. Tien, J. J. Lu, and L. Y. Jang, Phys. Rev. B \textbf{65}, 214416 (2002).
    
    \bibitem{Wenger} L. E. Wenger and P. H. Keesom, Phys. Rev. B \textbf{13}, 4053 (1976).
    
    \bibitem{Coey} J. M. D. Coey, S. von Molnar, and R. J. Gambino, Solid State 
    Commun. \textbf{24}, 167 (1977).
    
    \bibitem{Hattori} Y. Hattori, K. Fukamichi, K. Suzuki, H. Aruga-Katori, 
    and T. Goto, J. Phys.: Condens. Matter \textbf{7}, 4193 (1995).
    
    \bibitem{Kalvius} G. M. Kalvius, D. R. Noakes, and O. Hartmann, in \textit{Handbook 
    on the Physics and Chemistry of Rare Earths}, edited by K. A. Gschneidner, and 
    L. Eyring, (North Holland, Amsterdam, 2001), Vol. 32, p. 55.
    
    \bibitem{Dagotto} E. Dagotto, in \textit{Nanoscale Phase Separation and 
    Colossal Magnetoresistance}, (Springer-Verlag, Berlin, 2002); E. Dagotto, 
    cond-mat/0302550 and references therein.
    
    \bibitem{Mayr} M. Mayr, G. Alvarez, and E. Dagotto, Phys. Rev. B. \textbf{65}, 
    241202 (2002); G. Alvarez, and E. Dagotto, cond-mat/0305628.
    
    \bibitem{Millis} A. J. Millis, Phys. Rev. B \textbf{48}, 7183 (1993).
    
    \bibitem{Ishigaki} A. Ishigaki and T. Moriya, J. Phys. Soc. Jpn. \textbf{67}, 
    3924 (1998).
    
    \bibitem{Maple} M. B. Maple, M. C. de Andrade, J. Herrmann, Y. Dalichaouch, 
    D. A. Gajewski, C. L. Seaman, R. Chau, R. Movshovich, M. C. Aronson, and R. 
    Osborn, J. Low Temperature Phys. \textbf{99}, 223 (1995).
    
    \bibitem{Pfleiderer} C. Pfleiderer, D. Reznik, L. Pintschovius, H. von 
    L\"{o}hneysen, M. Garst, and A. Rosch, Nature \textbf{427}, 227 (2004).
    
    \bibitem{Bauer} E. Bauer, C. H. Booth, G. H. Kwei, R. Chau, and M. B. Maple, 
    Phys. Rev. B \textbf{65}, 245114 (2001).
    
    \bibitem{Booth} C. H. Booth, E. W. Scheidt, U. Killer, A. Weber, and S. 
    Kehrein, Phys. Rev. B \textbf{66}, 140402 (2002).
    
    \bibitem{Maksimov} I. Maksimov, F. J. Litterst, H. Rechenberg, M. A. C. de 
    Melo, R. Reyerhem, R. W. A. Hendrikx, T. J. Gortenmulder, J. A. Mydosh, and 
    S. S\"{u}llow, Phys. Rev. B \textbf{67}, 104405 (2003).
    
    \bibitem{Young} Ben-Li Young, D. E. MacLaughlin, M. S. Rose, K. Ishida, 
    O. O. Bernal, H. G. Lukefahr, K. Heuser, G. R. Stewart, N. P. Butch, P.-C. 
    Ho, and M. B. Maple, Phys. Rev. B \textbf{70}, 024401 (2004). 
    
\end{thebibliography}
\end{document}